\documentclass[journal=jacsat,manuscript=article]{achemso}

\usepackage[version=3]{mhchem} 
\usepackage{xcolor}
\usepackage{listings}
\usepackage{enumitem}
\usepackage{graphicx}
\usepackage{subcaption}
\usepackage{multicol}
\usepackage{float}
\usepackage{amsmath}
\usepackage{physics}
\usepackage{import}
\usepackage{longtable}
\usepackage{verbatim}
\usepackage{ulem} 
\usepackage{tikz}
\newcommand*\circled[1]{\tikz[baseline=(char.base)]{
            \node[shape=circle,draw,inner sep=2pt] (char) {#1};}}

\newcommand{\tocentryimage}[1]{
    \begin{center}
        \includegraphics[width=12cm]{#1} 
    \end{center}
}

\author{Jes\'us Lucia-Tamudo}
\affiliation[UR]{Institute of Chemistry and Pharmacy, University of Regensburg, Universitaetsstrasse 31, 93041 Regensburg, Germany}%
\alsoaffiliation{equal contribution}
\author{Michelle Menkel-Lantz}
\affiliation[CSULB]{Department of Chemistry and Biochemistry, California State University, Long Beach, 1250 Bellflower Boulevard, Long Beach, California 90840-9507, United States}%
\alsoaffiliation{equal contribution}
\author{Enrico Tapavicza}
\affiliation[CSULB]{Department of Chemistry and Biochemistry, California State University, Long Beach, 1250 Bellflower Boulevard, Long Beach, California 90840-9507, United States}%
\alsoaffiliation[UR]{Institute of Chemistry and Pharmacy, University of Regensburg, Universitaetsstrasse 31, 93041 Regensburg, Germany}%
\email{enrico.tapavicza@csulb.edu}

\title {First principles prediction of wavelength-dependent isomerization quantum yields of a second-generation molecular nanomotor}

\abbreviations{IR,NMR,UV}
\keywords{American Chemical Society, \LaTeX}

\begin{document}
\begin{tocentry}
\tocentryimage{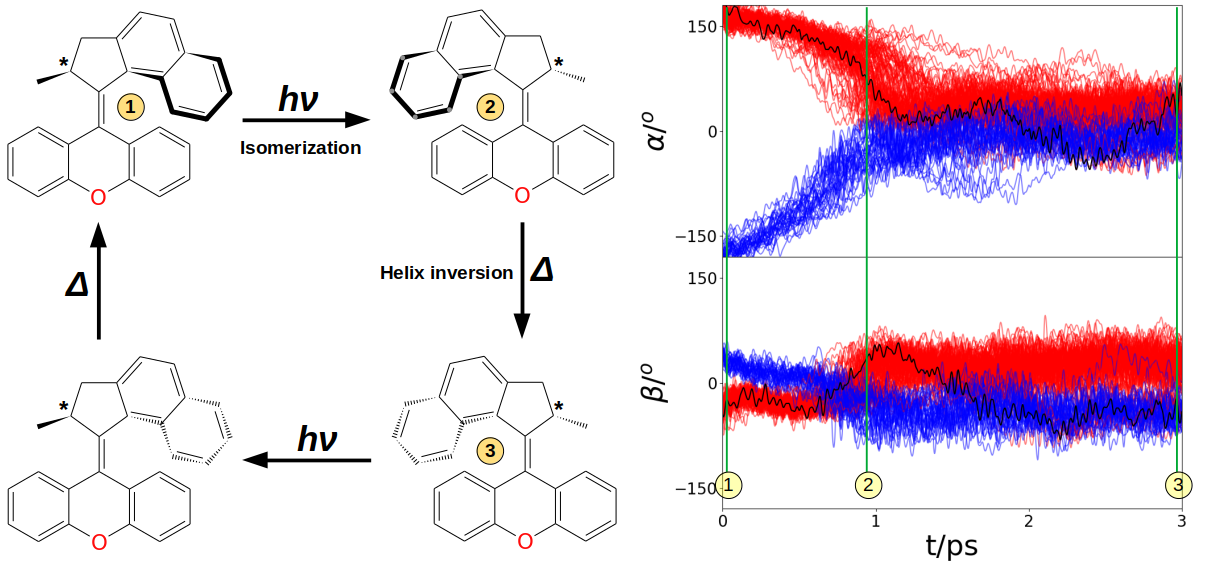}
\end{tocentry}

\begin{abstract}
  Second-generation molecular nanomotors are becoming more popular within the biomedical field and intense research is being conducted to increase their efficiency for light-induced ultrafast photoisomerization. A key requirement for designing efficient molecular nanomotors is ensuring unidirectional rotation during isomerization and thermal helix inversion. Here, we used non-adiabatic trajectory surface hopping molecular dynamics based on TDDFT to study the excited state dynamics of the stable M- and metastable P-conformers of a second-generation Feringa-type molecular nanomotor. From the trajectories, we computed quantum yields for clockwise and anti-clockwise photoisomerization. Results show that the helicity of the initial structure dictates the direction of the isomerization: 52 $\%$ of the trajectories starting from M-conformers isomerize clockwise, whereas 23 $\%$ of the trajectories starting from P-conformers isomerize anti-clockwise. The quantum yield for clockwise isomerization can be maximized by excitation in the center of the absorption spectrum (350 - 400 nm). In this region, the M-conformer exhibits its maximum absorption and maximum (clockwise) isomerization quantum yield, whereas the P-conformer shows negligible excitation probability and (anti-clockwise) isomerization quantum yield. Moreover, we also find several trajectories, where thermal helix inversion occurs in the hot ground state, directly after excited state relaxation. While helix inversion was observed to proceed in both directions, its net effect favors unidirectional rotation. We further report excited state lifetimes and details about the structural dynamics.
\end{abstract}

\section{Introduction}
\begin{figure}
    \centering
    \includegraphics[width=\textwidth]{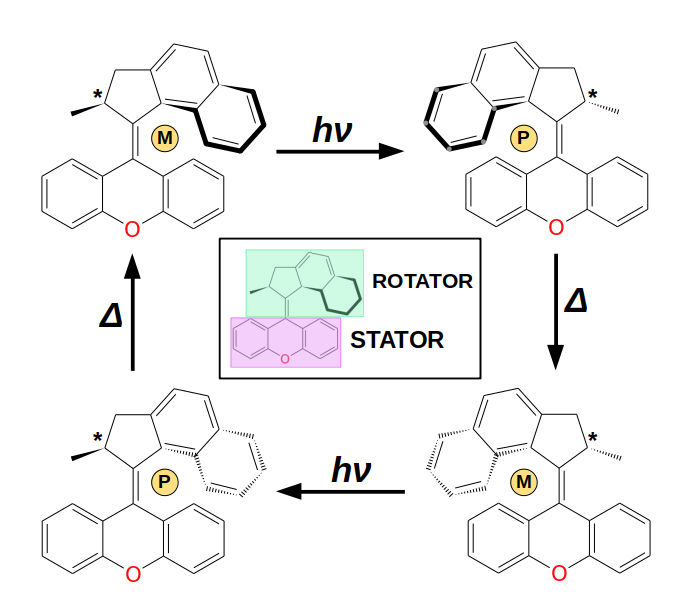}
    \caption{Graphical representation of the proposed stages of the unidirectional motion of CPNX. Helicity (M or P) is indicated. $h\nu$ indicate photochemical steps, whereas $\Delta$ indicate thermal reactions. The stereogenic center is marked with *.}
    \label{fig:rot}
\end{figure}

Light-driven molecular nanomotors (MNMs) hold significant potential for biomedical applications, including drug delivery, membrane permeation, and ion transport.\cite{garcia2019light,garcia2017molecular} Feringa-type overcrowded alkene motors consist of two polycyclic alkene-based subunits, usually referred to as the stator and rotator, connected by a carbon-carbon double bond (Figure \ref{fig:rot}). Ideally, light irradiation induces an E-Z photoisomerization\cite{stauch2016predicting} ($h\nu$ in Figure \ref{fig:rot}) followed by a thermal helix inversion ($\Delta$ in Figure \ref{fig:rot}).\cite{feringa1999,feringa2009} When these two elemental steps are repeated and occur in the same direction, one subunit completes a full 360º rotation relative to the other. Substantial research efforts, both experimental\cite{feringa1999,feringa_experimental,feringa2024,helix_inversion,rotary_spped,speed_and_helix,speed_and_unidirectionality} and computational,\cite{feringa2009,filatov_rotor,filatov_rotor_2,filatov_rotor_3,leticia_rotor,helicity_direction,marazzi2013} have been focused on understanding the dynamics of MNMs and on optimizing MNMs to increase their efficiency. With regard to the latter, studies focused on: (i) increasing photoisomerization quantum yield,\cite{filatov_rotor,feringa_experimental} (ii) increasing thermal helix inversion rates,\cite{helicity_direction,helix_inversion,speed_and_helix} (iii) increasing rotational speed,\cite{rotary_spped,speed_and_helix,speed_and_unidirectionality} determined by the time taken for isomerization and helix inversion, and (iv) ensuring unidirectional rotation.\cite{unidirectionality,marazzi2013,speed_and_unidirectionality} Computational studies focused on several aspects of the molecular mechanism and tuning factors of the process using a wide range of static and dynamic calculations,\cite{leticia_rotor,gonzalo} at different levels of theory such as time-dependent density functional theory (TDDFT)\cite{filatov_rotor_2}, its tight-binding version (TDDFTB)\cite{gonzalo}, TDDFT restricted ensemble Kohn-Sham (REKS),\cite{filatov_rotor,filatov_rotor_2} multiconfigurational methods,\cite{marazzi2013} or ADC(2)\cite{leticia_rotor}. This theoretical effort has led to increased understanding of the mechanism of MNMs and which factors influence its functioning. For instance, it has been shown that an increasing polarity of the solvent increases the excited state lifetime of the MNM, while increased viscosity produces the opposite effect.\cite{feringa2009} González and co-workers performed computational excited-state dynamics of a molecular rotor embedded in a solvent (DMSO), gauging the influence of the solvent on the MNM photoisomerization.\cite{leticia_rotor} Furthermore, it has been found that hydrogen bonds between MNM and solvent decrease the rotation efficiency.\cite{filatov_rotor_3} Also substituents play an important role in the efficiency of the photoinduced rotation: the gap between the heterolytic and homolytic partial bond breaking of the central carbon-carbon double bond should be minimal to lead to a faster photoisomerisation and enhanced quantum yield.\cite{filatov_rotor_2} Moreover, a stator that decreases steric repulsion in the fjord region also improves the photoisomerization rate. On the other hand, increasing the size of substituents at the stereogenic center accelerates the rotary motion because it reduces the energetic barrier for thermal helix inversion.\cite{speed_and_unidirectionality}
Despite the gained understanding of the working mechanism, there are still various aspects that require improvement of molecular properties. 
For instance, for application in biological tissue, activation by near-infrared instead of UV/Vis light would be beneficial, as photons in this region exhibit deeper penetration depth in biological tissues.\cite{ash2017effect} Using two-photon excitation, near-infrared activation has recently been achieved \cite{liu2019near}, but it still requires high photon intensity, due to the low two-photon absorption cross sections. Another property to be increased is high photostability, to ensure the MNM can sustain as many cycles as possible without degradation.\cite{degradation}

Feringa and co-workers designed a pioneering MNM that served as a blueprint for the development of first-generation MNMs, followed by second and third-generation versions.\cite{feringa1999} Unidirectional rotation is essentially a consequence of the the chiral structure of the MNM.\cite{speed_and_unidirectionality} While first-generation MNMs feature two stereogenic centers, second-generation MNMs contain only one, and third-generation MNMs are formed by merging two enantiomers from second-generation MNMs to create a pseudo-asymmetric center responsible for unidirectional rotation. The reduction in stereogenic centers has facilitated the synthesis of second- and third-generation MNMs, simplifying the process compared to their predecessors.\cite{pooler2021designing}

During 360º rotation, the MNM adopts several conformers (Figure \ref{fig:rot}). Broadly, these can be classified into stable and metastable conformers, with opposite rotor helicity. While the absolute stereochemical configuration of the MNM is arbitrary, we use here the absolute configuration marked with * in Figure \ref{fig:rot}, in which M-conformers (left-handed helicity) are stable, whereas P-conformers (right-handed helicity) are metastable. To ensure unidirectional rotation, both photoinduced isomerizations as well as thermal helix inversion should ideally proceed in the same direction; rotation in the opposite direction should not occur. 
While at thermal equilibrium, typically the stable conformer (the M-helical conformer in our example, Figure \ref{fig:rot}) dominates the equilibrium of structures, this is not the case in non-equilibrium situations, such as the hot ground state after photorelaxation, as the primarily produced conformer is typically the metastable conformer. In this case, both left- and right-handed helicity conformers may coexist and both conformers are subject to photoexcitation in the next excitation cycle. 
The coexistence of several conformers could compromise unidirectional rotation, if both stable and metastable conformers undergo photoisomerization in opposite directions, as previously hypothesized.\cite{helicity_direction,filatov_rotor} Most computational studies that focussed on the photochemical isomerization have examined the primary isomerization of the stable conformer to the metastable conformer.\cite{leticia_rotor,filatov_rotor,filatov_rotor_2,filatov_rotor_3,gonzalo} On the contrary, few research has been conducted to get insights into the excited-state relaxation of the metastable conformer (in our case P-conformer).\cite{speed_and_unidirectionality,helicity_direction}
 To address the different photochemical reactivity of the stable and metastable conformers, we investigate the excited-state decay pathways of both M- and P-conformers upon photoexcitation into the first excited singlet state (S$_1$). Specifically, we aim at verifying the hypothesis that conformers with opposite helicity undergo photoisomerization in different rotational directions.\cite{helicity_direction,filatov_rotor} Furthermore, we aim at determining the conditions at which the unidirectional rotation can be maximized. Using fewest switches surface hopping\cite{Tully1990} based on TDDFT (TDDFT-SH), we study the dynamics of photoisomerization mechanism with a focus on predicting the quantum yield of clockwise and anti-clockwise photoisomerization, and its dependency on the excitation wavelength of the recently synthesized 9-(2’-methyl-2’,3’-dihydro-1’H-cyclopenta[a]naphthalen-1’-ylidene)-9H-xanthene (CPNX),\cite{augulis2009light} depicted in Figure \ref{fig:rot}. 

\section{Computational Details}
\subsection{Geometry Optimization}
Initially, we searched the conformational space of structures of CPNX using the software CREST,\cite{crest} employing metadynamics\cite{laio2002escaping} simulations with extended tight-binding method (XTB)\cite{xtb} in the gas phase. Structures within an energy window of 10 kcal/mol with respect to the most stable structure found were further optimized using density functional theory (DFT)\cite{Haeser1989} at the PBE0/def2-SVP level of theory\cite{pbe,pbe0,def2_1,def2_2,def2_3,def2_4} to locate the local minima of each species. Frequency calculations\cite{Deglmann2002,Deglmann2002b} were performed at the same level of theory to confirm that these structures corresponded to minima on the potential energy surface. The relative population of each conformer was assessed based on a Boltzmann distribution at 300K. To verify the accuracy of the DFT results, the total energy using second-order approximate coupled cluster (CC2)/def2-TZVP was computed.\cite{cc2,def2_1,def2_2,def2_3,def2_4}
All DFT and CC2 calculations were performed with TURBOMOLE
\cite{TURBOMOLE,balasubramani2020turbomole,franzke2023turbomole}
and employed default convergence criteria as provided by the program.

\subsection{Generation of ground state ensemble of structures}

\subsubsection{Replica Exchange Molecular Dynamics}
Five replica exchange molecular dynamics (REMD)\cite{Sugita1999} simulations using XTB were conducted in order to sample the conformational space of CPNX.\cite{xtb} For each conformer, eight independent trajectories were generated at different temperatures (300, 350, 400, 450, 500, 550, 600, and 650 K). A time-step of 40 au was used to propagate the nuclear degrees of freedom. Every 200 MD steps, the probability of switching between two trajectories at neighboring temperatures was evaluated. A comparison of this probability with a random number determined whether the current structures of the two trajectories would be exchanged. Further implementation details for REMD can be found elsewhere.\cite{xtb,Cisneros2017,tapavicza2022conformationally} The total simulation time for each trajectory amounted to 9.5 ns, leading to a combined total of 47.5 ns, which yielded a ground-state ensemble of geometries consisting of approximately 50,000 structures. REMD was carried out with a home-written script using TURBOMOLE.\cite{TURBOMOLE,balasubramani2020turbomole,franzke2023turbomole}

\subsubsection{Clustering of the structures}
The structures from the REMD trajectories were clustered using the density-based spatial clustering of applications with noise (DBSCAN) algorithm, based on the structural differences between structures, as implemented in AmberTools 22.\cite{amber,clustering,dbscan} For this algorithm, the user defines the minimum number of points that constitute a cluster and the neighborhood density ($\varepsilon$) around a point to determine whether it can form part of a cluster: it defines the maximum distance between two points to be considered neighbors. After selecting these parameters appropriately, the algorithm optimizes the number of clusters. In this study, $\varepsilon$ was set to 0.5 \AA{}, and the minimum number of points required to form a cluster was set to 1. Subsequently, the 372 starting structures for TDDFT trajectory surface hopping (TDDFT-TSH) (see below) were clustered using the same methodology to verify whether this reduced ensemble was representative of the complete REMD ensemble.

\subsection{Computation of absorption spectra}
The lowest 5 electronic excitation energies and oscillator strengths were computed for the 372 selected geometries using CC2/def2-TZVP\cite{cc2,def2_1,def2_2,def2_3,def2_4} and the resolution of identity, as implemented in TURBOMOLE. \cite{Hattig02,balasubramani2020turbomole,franzke2023turbomole} Absorption spectra were generated as a stick spectrum of the five lowest excitation energies and oscillator strengths broadened by Gaussian functions with a fixed full width at half maximum of 0.1 eV and converted to molar decadic extinction coefficients, as previously described \cite{Epstein2013,grathwol2019azologization,tapavicza2021elucidating}.

\subsection{Surface Hopping Dynamics}
375 structures, selected at equal distances from the REMD ensemble, were used as starting structures for TDDFT-SH dynamics.\cite{Tapavicza2011,Tapavicza2013} Three trajectories showed numerical errors along the excited-state dynamics and were discarded from the study. As a result, 372 trajectories are considered in this work. TDDFT-SH dynamics simulations were initiated in the first excited singlet state (S$_1$) and were propagated for 3 ps employing the hybrid PBE0 exchange-correlation functional\cite{pbe,pbe0} and the def2-SVP basis set,\cite{def2_1,def2_2,def2_3,def2_4} as implemented\cite{Furche2002} in TURBOMOLE V7.2.\cite{TURBOMOLE,balasubramani2020turbomole} TDDFT was used within the Tamm-Dancoff approximation (TDA) to minimize instabilities close to conical intersections.\cite{tamm_dancoff,Levine2006,Tapavicza2008} A total of two excited states were computed. A time-step of 40 au was used to propagate the nuclear degrees of freedom.
TDDFT-SH simulations employed analytic non-adiabatic coupling (NAC) vectors\cite{Tavernelli2009, Tavernelli2009a, Send2010}, which allow proper velocity rescaling after surface hops, as proposed by Tully\cite{Tully1990}. The hopping probability between excited states was evaluated at each time step, but in case of negative excitation energies, which occur at conical intersections within TDA,\cite{Levine2006,Tapavicza2008} a transition to the ground state was forced.

\subsection{Wavelength-dependent product quantum yields}
Wavelength-dependent product quantum yields were calculated as previously described.\cite{Thompson2018,Tapavicza2018} To this end, we clustered the trajectories into {\it successful} and {\it unsuccessful} trajectories, depending on whether they underwent complete Z-E isomerization around their central double bond (C9=C1', defined in Figure \ref{fig:dihed}) or whether they decayed to the ground state unreactively. We further distinguished between clockwise and anti-clockwise isomerization. The wavelength-dependent product quantum yield ($\Phi_j(\lambda)$) for the isomerization of conformer $j$ was then obtained by dividing the absorption spectrum averaged over the initial structures of the corresponding successful trajectories by the average absorption spectrum of the initial structures of all trajectories, according to
\begin{equation}
    \Phi_j(\lambda)=\frac{\sum_i ^{N_{succ,j}}\sigma_i(\lambda)}{\sum_i ^{N_{total,j}}\sigma_i(\lambda)}, 
    \label{eq1}
\end{equation}
where $\sigma_i(\lambda)$ indicates the absorption spectrum of the initial structure with a conformation $j$ of trajectory $i$, and $N_{succ,j}$ and $N_{total,j}$ indicate the number of successful trajectories and total number of trajectories, respectively. Spectra were calculated using CC2 as described above.

\section{Results \& discussion}
\subsection{Ground-state equilibrium structures}
\begin{figure}
    \centering
    \includegraphics[width=0.8\textwidth]{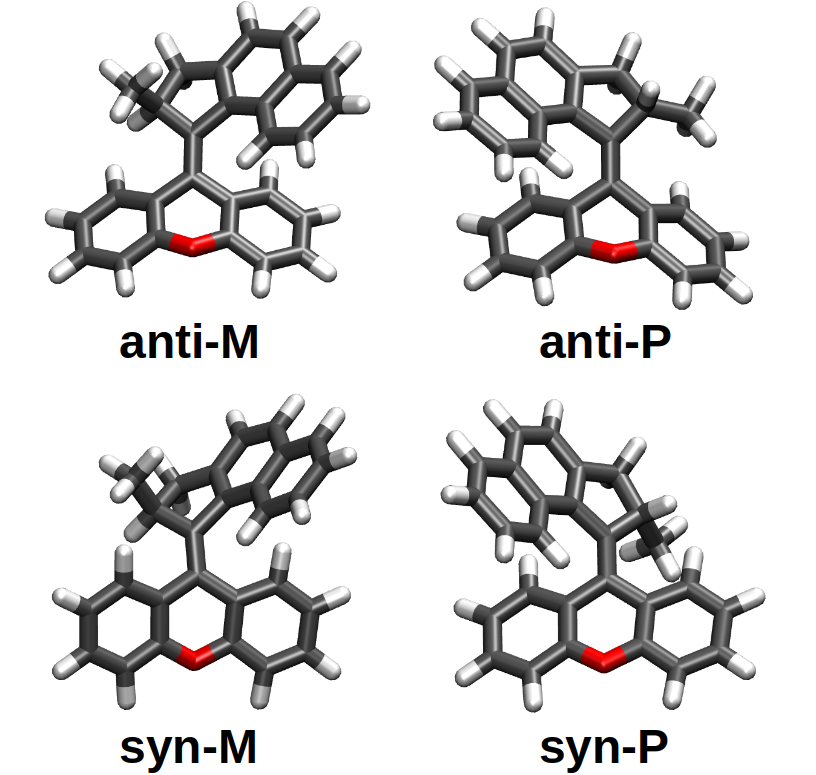}
    \caption{Graphical representation of the different equilibrium conformers of CPNX optimized using PBE0/def2-SVP. Color code: gray for carbon, white for hydrogen and red for oxygen.}
    \label{fig:iso}
\end{figure}

The XTB-level conformational search with CREST yields seven distinct conformers. Each conformer was further optimized using PBE0/def2-SVP (Cartesian coordinates of the structures, labeled I1-I7 are given in section 1 of the Supporting Information (SI)). Based on the similarity of the dihedral angles (Table S1 in the SI) and total energies (Table S2 in the SI), we further reduced the number of conformers to four: anti-M, syn-M, anti-P, and syn-P (Figure \ref{fig:iso}, Table \ref{tbl:pop}). The notation for these conformers is as follows: M-helical conformers are characterized by a left-handed helicity, while P-helical conformers exhibit a right-handed helicity. The anti-folded or syn-folded character is determined by the orientation of the stator and rotator relative to the olefinic plane (containing the C9=C1' double bond). When both subunits are oriented in the same direction relative to the olefinic plane, the conformation is syn-folded; otherwise, it is anti-folded. I1 and I2 correspond to the same anti-M structure whereas conformers I5 and I6 are associated with the syn-M isomer. In the case of isomers I3 and I4, both represent a typical syn-P structure, although the methyl group points towards a slightly different direction and thus, their geometrical parameters and relative energies differ. The most stable structure for each conformer was used as the representative for anti-M (I1), syn-M (I5), anti-P (I7), and syn-P (I3).

Populations obtained from total energies computed by CC2/def2-TZVP (Table S2 in the SI) and PBE0/def2-SVP (Table \ref{tbl:pop}) are qualitatively similar. The anti-M structure is the most stable conformer, with a population of $99.51\%$ at 300K (Table \ref{tbl:pop}). The syn-M and syn-P conformers have similar populations, $0.36\%$ and $0.13\%$, respectively. Finally, anti-P shows negligible population, $\sim0.00002\%$. These data are consistent with the relative stability for the conformers of the analogous sulfur-containing MNM studied by Durbeej \textit{et al.}\cite{s_rotor}

\subsection{Analysis and clustering of REMD trajectories}
All structures from the five REMD simulations were clustered using the DBSCAN algorithm.\cite{dbscan} Four different clusters were identified and labeled in terms of the cluster centroid. These centroids coincide with the four conformers found within the conformational search (Table \ref{tbl:pop}). 

We notice differences in the populations of the REMD ensemble and the Boltzmann probabilities of the optimized structures (Table \ref{tbl:pop}). The large population differences indicate that REMD did not converge because the number of transitions between structures of REMD trajectories at different temperatures was not enough to ensure proper capturing of the statistics of the different conformers, as conformers were trapped for long simulation periods in one local minimum. 

From the REMD ensemble, 372 samples were taken as initial structures for the TDDFT-SH simulations (REMD sub-ensemble). The sub-ensemble has a similar conformer distribution as the full REMD ensemble (Table \ref{tbl:pop}).%

\begin{table}[htb!]
  \centering
  \caption{Statistical distribution of the CPNX conformers in terms of a Boltzmann distribution using PBE0/def2-SVP for the equilibrium structures, and from the ensemble (and representative sub-ensemble) of geometries obtained from REMD simulations. Boltzmann populations arise from the addition of the individual populations of those structures that represent the same conformer (Table S2 in the Supporting Information).}
  \label{tbl:pop}
  \begin{tabular}{cccc}
    \hline
    Conformer & Boltzmann Population & Population & Population  \\
    & equilibrium &  REMD &  REMD sub-ensemble  \\
    & structures &   (all structures)&   (372 structures)  \\
    \hline
    \textit{anti}-(M) & 0.9951 & 0.5460 & 0.5780 \\
    \textit{syn}-(M)  & $3.630 \cdot 10^{-3}$ & $1.987 \cdot 10^{-3}$ & $1.882 \cdot 10^{-2}$ \\
    \textit{anti}-(P) & $1.940 \cdot 10^{-7}$ & 0.1770 & 0.1425 \\
    \textit{syn}-(P)  & $1.310 \cdot 10^{-3}$ & 0.2750 & 0.2608 \\
    \hline
  \end{tabular}
\end{table}

\subsubsection{Geometric parameters}

\begin{figure}
    \centering
    \includegraphics[width=\textwidth]{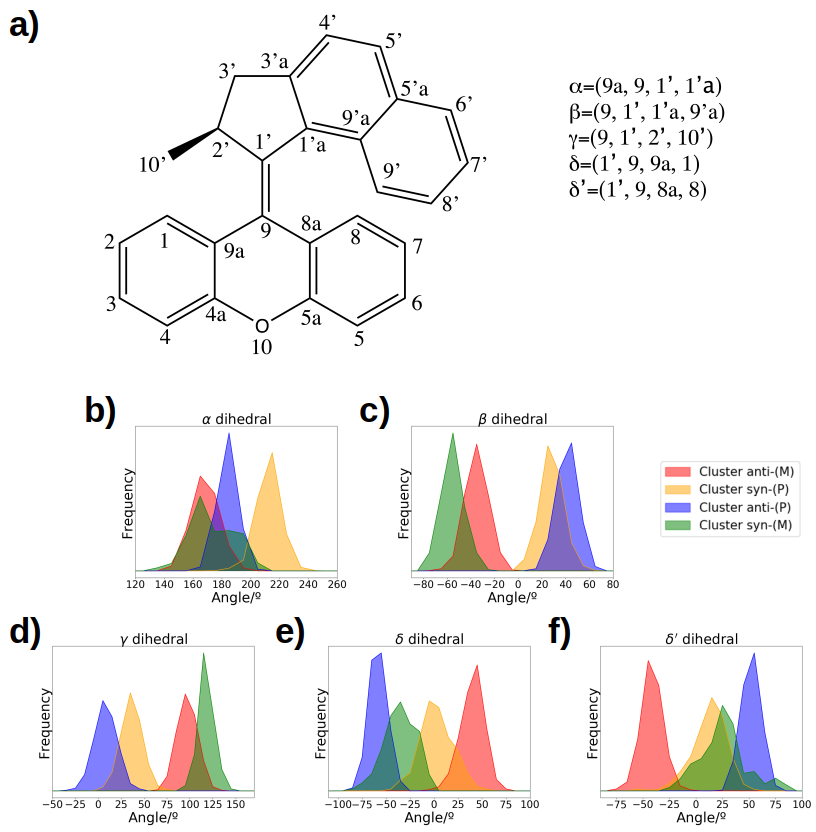}
    \caption{a) Lewis structure of CPNX and definition of characteristic dihedral angles. Distributions of dihedral angles of the structures of the REMD simulations color coded for each cluster are show in panels b)--f), as labeled.}
    \label{fig:dihed}
\end{figure}

We analyzed the conformers in terms of five dihedral angles, defined in Figure \ref{fig:dihed}a. The dihedral angle distributions (Figure \ref{fig:dihed}b-f) are consistent with the values of the equilibrium structures (Table S1 in the Supporting Information) and can be used to identify and distinguish the different conformers. The distributions reveal that the helicity is associated with the sign of the dihedral angle $\beta$ (Figure \ref{fig:dihed}c): Negative $\beta$ values indicate M-helicity, whereas positive $\beta$ values correspond to P-helicity. 

anti-M and syn-M conformers can be distinguished by their distinct $\delta$- and $\delta'$-distributions, Figure \ref{fig:dihed}e and \ref{fig:dihed}f, respectively.
However, small overlaps between the two distributions exist; thus for a few cases, the angle distributions are not sufficient to distinguish the two conformers and more geometric parameters have to be considered.

syn-P and anti-P conformers can also be distinguished based on their distinct $\delta$- and $\delta'$-distributions, but as in the case of the M-conformers, small overlapping regions between the distributions of the two P-conformers exist. 

\subsection{Absorption Spectra}
Analysis of the CC2 results of the lowest five electronic excited states of the equilibrium geometries (Table \ref{tbl:eeandosc}) shows that, for all the conformers, S$_1$ is bright with oscillator strengths at least one magnitude higher than those of S$_2$-S$_5$. For all conformers, S$_1$ has a HOMO-LUMO character, corresponding to a $\pi-\pi^*$ transition (orbitals and transition densities shown in Figures S1-S4 in the SI). S$_1$ excitation energies range between 2.74-3.75 eV, mainly located in the UV region of the electromagnetic spectrum, with syn-P being the only one that has its major band in the visible region.

\begin{table}[htb!]
  \centering
  \caption{CC2 excitation energies in eV, oscillator strengths in au (in parentheses), and dominant orbital contribution for the five lowest excited states of ground state equilibrium geometries of all four conformers optimized with PBE0/def2-SVP. H and L denote highest occupied molecular orbital (HOMO) and lowest unoccupied molecular orbital (LUMO), respectively.}
  \label{tbl:eeandosc}
  \small
  \begin{tabular}{lrrrrr}
    \hline \hline
    \multicolumn{1}{l}{} & \multicolumn{1}{l}{S$_1$} & \multicolumn{1}{l}{S$_2$} & \multicolumn{1}{l}{S$_3$}& \multicolumn{1}{l}{S$_4$} & \multicolumn{1}{l}{S$_5$}   \\
        \hline
         anti-M &   3.428 & 4.056 & 4.187 & 4.344 & 4.562 \\
           &   (0.4463) & (0.002901) & (0.04247) & (0.02397) & (0.005384) \\
           &   \multicolumn{1}{l}{ H $\xrightarrow[]{}$ L ($95\%$)} & \multicolumn{1}{l}{H $\xrightarrow[]{}$ L+1 ($52\%$)} & \multicolumn{1}{l}{H-1 $\xrightarrow[]{}$ L ($33\%$)} & \multicolumn{1}{l}{H $\xrightarrow[]{}$ L+1 ($36\%$)} & \multicolumn{1}{l}{H-3 $\xrightarrow[]{}$ L ($38\%$)} \\
           &    & \multicolumn{1}{l}{H-1 $\xrightarrow[]{}$ L ($34\%$)} & \multicolumn{1}{l}{H $\xrightarrow[]{}$ L+1 ($26\%$)} & \multicolumn{1}{l}{H-1 $\xrightarrow[]{}$ L ($28\%$)} & \multicolumn{1}{l}{H $\xrightarrow[]{}$ L+1 ($19\%$)} \\
        \hline
         anti-P   & 3.755 & 4.140 & 4.299 & 4.515 & 4.717 \\
          &   (0.3662) & (0.01518) & (0.04128) & (0.01377) & (0.03392) \\
          &   \multicolumn{1}{l}{H $\xrightarrow[]{}$ L ($85\%$)} & \multicolumn{1}{l}{H $\xrightarrow[]{}$ L+1 ($31\%$)} & \multicolumn{1}{l}{H-1 $\xrightarrow[]{}$ L ($36\%$)} & \multicolumn{1}{l}{H $\xrightarrow[]{}$ L+1 ($44\%$)} & \multicolumn{1}{l}{H-3 $\xrightarrow[]{}$ L ($34\%$)} \\
            &  & \multicolumn{1}{l}{H $\xrightarrow[]{}$ L+2 ($28\%$)} & \multicolumn{1}{l}{H $\xrightarrow[]{}$ L+1 ($19\%$)} & \multicolumn{1}{l}{H-1 $\xrightarrow[]{}$ L ($23\%$)} & \multicolumn{1}{l}{H $\xrightarrow[]{}$ L+1 ($20\%$)} \\
        \hline
         syn-M   & 3.381 & 4.011 & 4.053 & 4.222 & 4.346 \\
          &   (0.4517) & (0.005296) & (0.009919) & (0.07164) & (0.0009657) \\
          &   \multicolumn{1}{l}{H $\xrightarrow[]{}$ L ($91\%$)} & \multicolumn{1}{l}{H $\xrightarrow[]{}$ L+1 ($30\%$)} & \multicolumn{1}{l}{H $\xrightarrow[]{}$ L+1 ($53\%$)} & \multicolumn{1}{l}{H-1 $\xrightarrow[]{}$ L ($54\%$)} & \multicolumn{1}{l}{H $\xrightarrow[]{}$ L+3 ($78\%$)} \\
          &    & \multicolumn{1}{l}{H-3 $\xrightarrow[]{}$ L ($17\%$)} & \multicolumn{1}{l}{H $\xrightarrow[]{}$ L+2 ($15\%$)} & \multicolumn{1}{l}{H $\xrightarrow[]{}$ L+1 ($17\%$)} & \multicolumn{1}{l}{H-1 $\xrightarrow[]{}$ L+3 ($7\%$)} \\
        \hline
         syn-P   & 2.745 & 3.778 & 3.854 & 3.919 & 3.973 \\
          &   (0.4410) & (0.004166) & (0.008005) & (0.03583) & (0.002100) \\
          &   \multicolumn{1}{l}{H $\xrightarrow[]{}$ L ($93\%$)} & \multicolumn{1}{l}{H $\xrightarrow[]{}$ L+1 ($63\%$)} & \multicolumn{1}{l}{H $\xrightarrow[]{}$ L+2 ($29\%$)} & \multicolumn{1}{l}{H $\xrightarrow[]{}$ L+3 ($33\%$)} & \multicolumn{1}{l}{H $\xrightarrow[]{}$ L+3 ($50\%$)} \\
          &    & \multicolumn{1}{l}{H-3 $\xrightarrow[]{}$ L ($13\%$)} & \multicolumn{1}{l}{H-3 $\xrightarrow[]{}$ L ($19\%$)} & \multicolumn{1}{l}{H $\xrightarrow[]{}$ L+2 ($26\%$)} & \multicolumn{1}{l}{H-3 $\xrightarrow[]{}$ L ($12\%$)} \\
         \hline \hline
    \end{tabular}
\end{table}

\begin{figure}
    \centering
    \includegraphics[width=\textwidth]{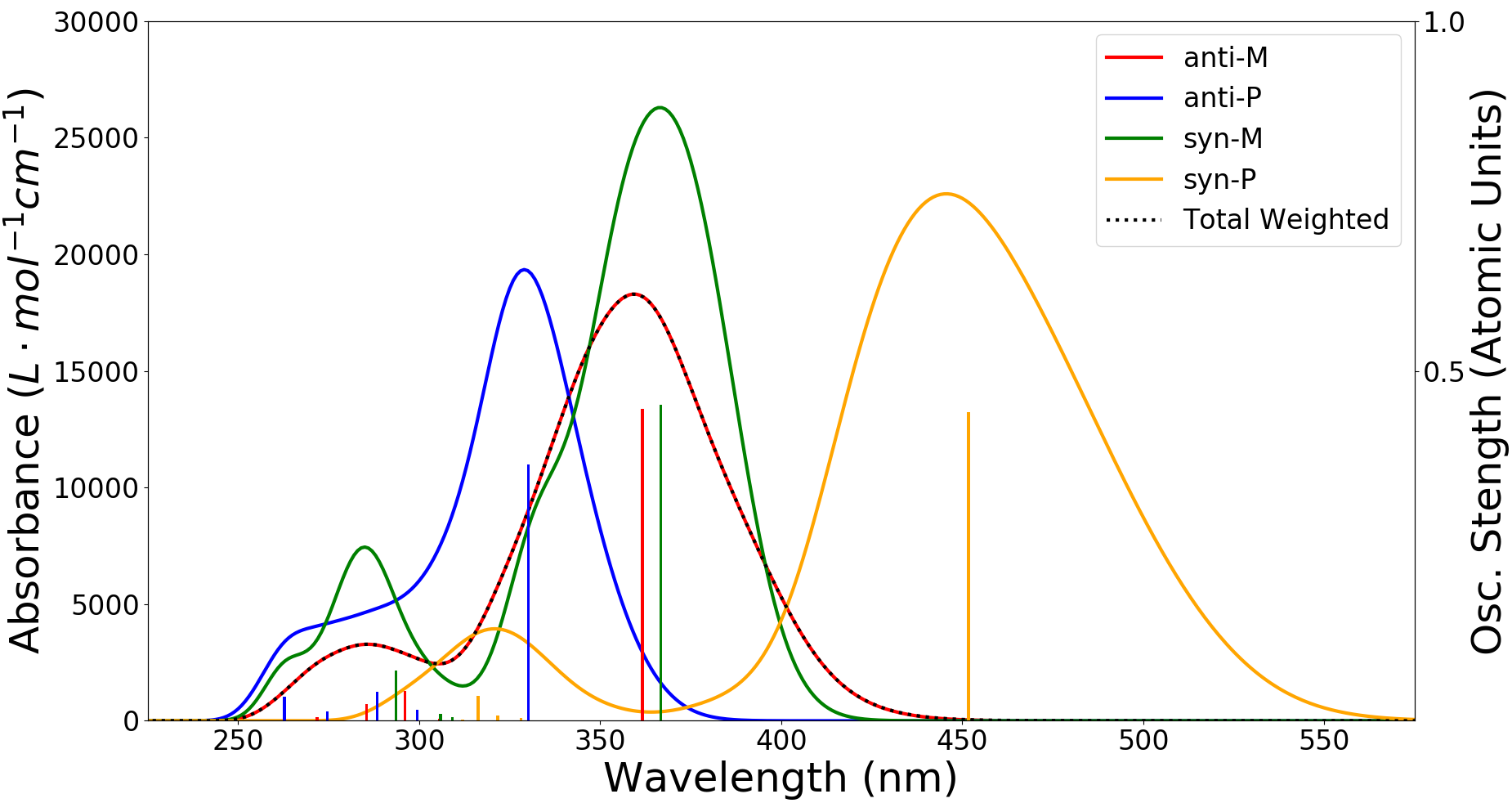}
    \caption{Absorption spectra for each conformer (solid lines, colors as indicated) computed from the REMD sub-ensemble (Table \ref{tbl:tsh}). For the total spectrum at thermal equilibrium (black), each conformer spectrum was weighted by the corresponding Boltzmann population (Table \ref{tbl:pop}). Stick spectra show excitation energies and oscillator strengths (length representation) of the equilibrium structures (Figure \ref{fig:iso}, Table \ref{tbl:eeandosc}), colors as indicated.}
    \label{fig:abs}
\end{figure}

In addition, the theoretical absorption spectrum of each conformer $\sigma_{\text{conf}}(\lambda)$ was calculated as the average of the spectra of individual structures of the REMD sub-ensemble (Figure \ref{fig:abs}) according to 
\begin{equation}
    \sigma_{\text{conf}}(\lambda)=\frac{1}{N_{\text{conf}}}\sum_i \sigma_{i},(\lambda)
\end{equation}
where $N_{\text{conf}}$ is the number of structures for a given conformer and $\sigma_i(\lambda)$ denotes the absorption spectrum of one individual structure.
The total absorption spectrum at thermal equilibrium was computed by weighting each conformer's spectrum according to its Boltzmann probability at 300 K. Since anti-M accounts for $99.51\%$ of the structures (Table \ref{tbl:pop}), the total spectrum is practically identical to that of anti-M. 

The S$_1$ excitation energies of the equilibrium geometries (stick spectrum in Figure \ref{fig:abs}) align closely with the peak positions of the maximum absorption bands of each conformer. The spectra of anti-M and syn-M are very similar to each other. In contrast, the two P-conformers exhibit very different absorption spectra: anti-P has its main absorption band blue-shifted relative to the two M-conformer bands, whereas the main absorption band of the syn-P conformer appears on the red side of the spectrum. We conclude that in the window between $360-400$ nm, M-conformers can be predominantly excited.

\subsection{TDDFT-SH dynamics}
\begin{table}[htb!]
  \centering
  \caption{Conformer distribution of TDDFT-SH the initial structures and number of trajectories undergoing clockwise and anti-clockwise photoisomerization. Unsuccessful trajectories did not undergo Z-E isomerization within 3 ps simulation time.}
  \label{tbl:tsh}
  \begin{tabular}{ccccc}
    \hline
    Conformer & Initial structures & Clockwise&  Anti-clockwise &Unsuccessful \\
    \hline
    anti-(M) & 215 & 114 &  0&101\\
    syn-(M) & 7 & 3 &  0&4\\
    anti-(P) & 53 & 0&  30&23\\
    syn-(P) & 97 & 0&  4&93\\
    \hline
  \end{tabular} 
\end{table}

\begin{figure}
    \centering
    \includegraphics[width=\textwidth]{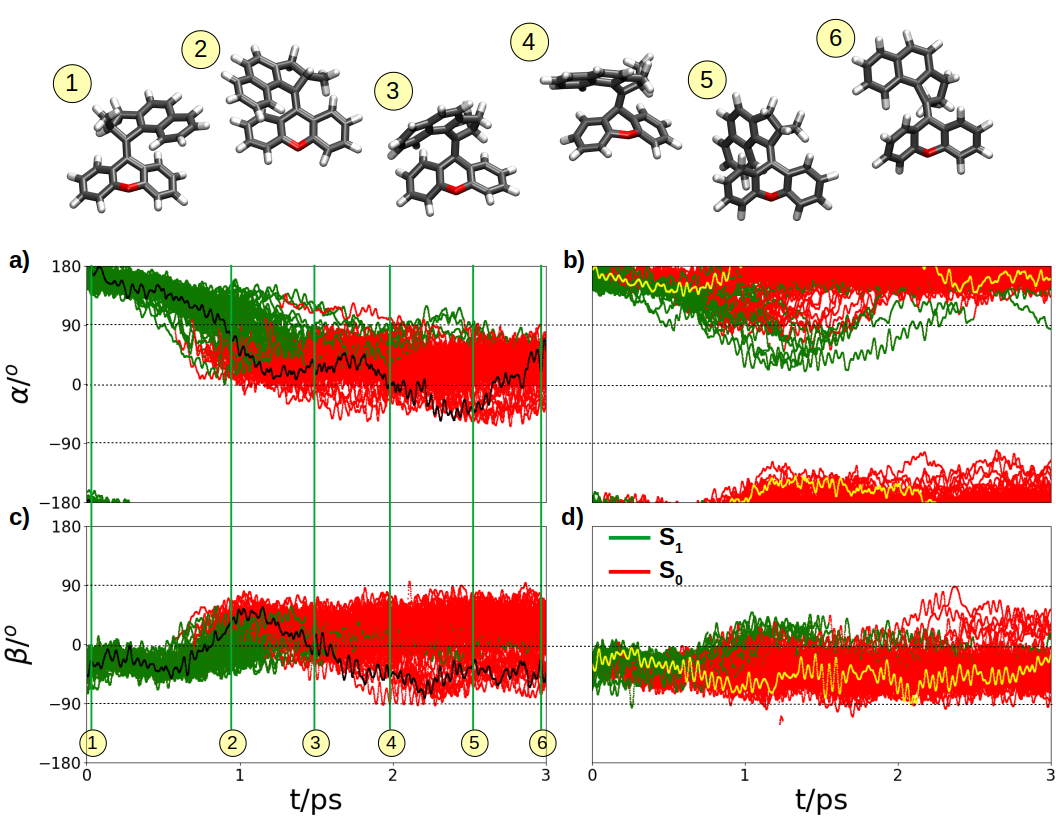}
    \caption{Evolution of dihedral angles $\alpha$ (a, b) and $\beta$ (c, d) along the trajectories with initial M-conformers. Panels on the left (a, c) represent trajectories that isomerize, whereas panels on the right (b, d) account for trajectories without isomerization. Green and red represent the electronic state (S$_1$ or S$_0$, respectively) of the molecule along the trajectory. The black lines (left panels) belong to an example trajectory that undergoes thermal helix inversion in the hot ground state after E-Z isomerization; associated structures for this trajectory are labeled with numbers from 1 to 6; potential energies along the trajectory are shown in Figure \ref{fig:pes}a. Yellow lines (right panels) correspond to the example trajectory displayed in Figure \ref{fig:pes}b.}
    \label{fig:dihe_time_m}
\end{figure}

\begin{figure}
    \centering
    \includegraphics[width=\textwidth]{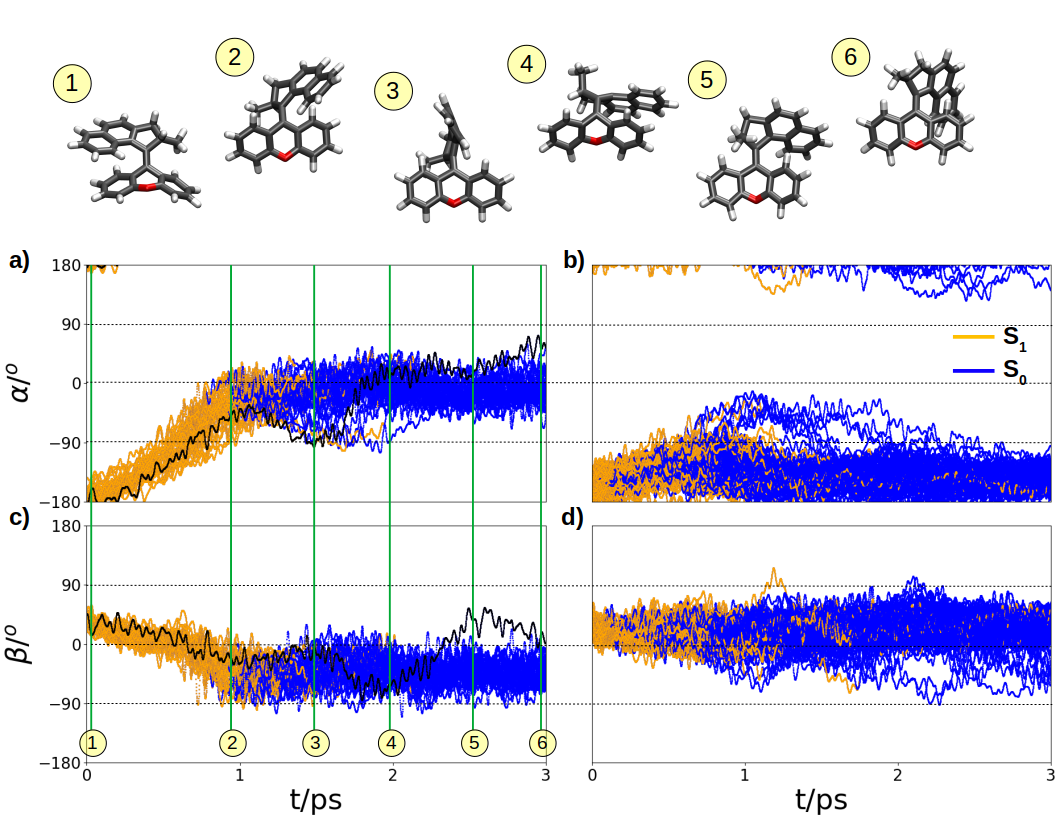}
    \caption{Evolution of dihedral angles $\alpha$ (a, b) and $\beta$ (c, d) along the trajectories with initial M-conformers. Panels on the left (a, c) represent trajectories that isomerize, whereas panels on the right (b, d) account for trajectories without isomerization. Orange and blue represent the electronic state (S$_1$ or S$_0$, respectively) of the molecule along the trajectory. The black lines (left panels) belong to an example trajectory that undergoes thermal helix inversion in the ground state after E-Z isomerization; associated structures for this trajectory are labeled with numbers from 1 to 6.}
    \label{fig:dihe_time_p}
\end{figure}

Within the first 3 ps of the TDDFT-SH simulations, $97.8 \%$ of initial M-conformers and $86.7 \%$ of initial P-conformers decayed to S$_0$. The results of the simulations of the 372 selected structures (Table \ref{tbl:tsh}) showed that out of the 222 trajectories starting from M-helical initial structures, 117 (52 \%) undergo clockwise photoisomerization (Figure \ref{fig:dihe_time_m}a and \ref{fig:dihe_time_m}c), while 105 trajectories do not show any photochemical transformation within 3 ps (Figure \ref{fig:dihe_time_m}b and \ref{fig:dihe_time_m}d). From the 150 trajectories starting from P-helical structures, 34 (23 \%) isomerize anti-clockwise (Figure \ref{fig:dihe_time_p}a and \ref{fig:dihe_time_p}c), while 116 trajectories do not show any E-Z isomerization within 3 ps (Figure \ref{fig:dihe_time_p}b and \ref{fig:dihe_time_p}d). None of the initial structures with M-helicity underwent anti-clockwise isomerization, and none of the initial structures with P-helicity underwent clockwise isomerization.


\begin{figure}
    \centering
    \includegraphics[width=\textwidth]{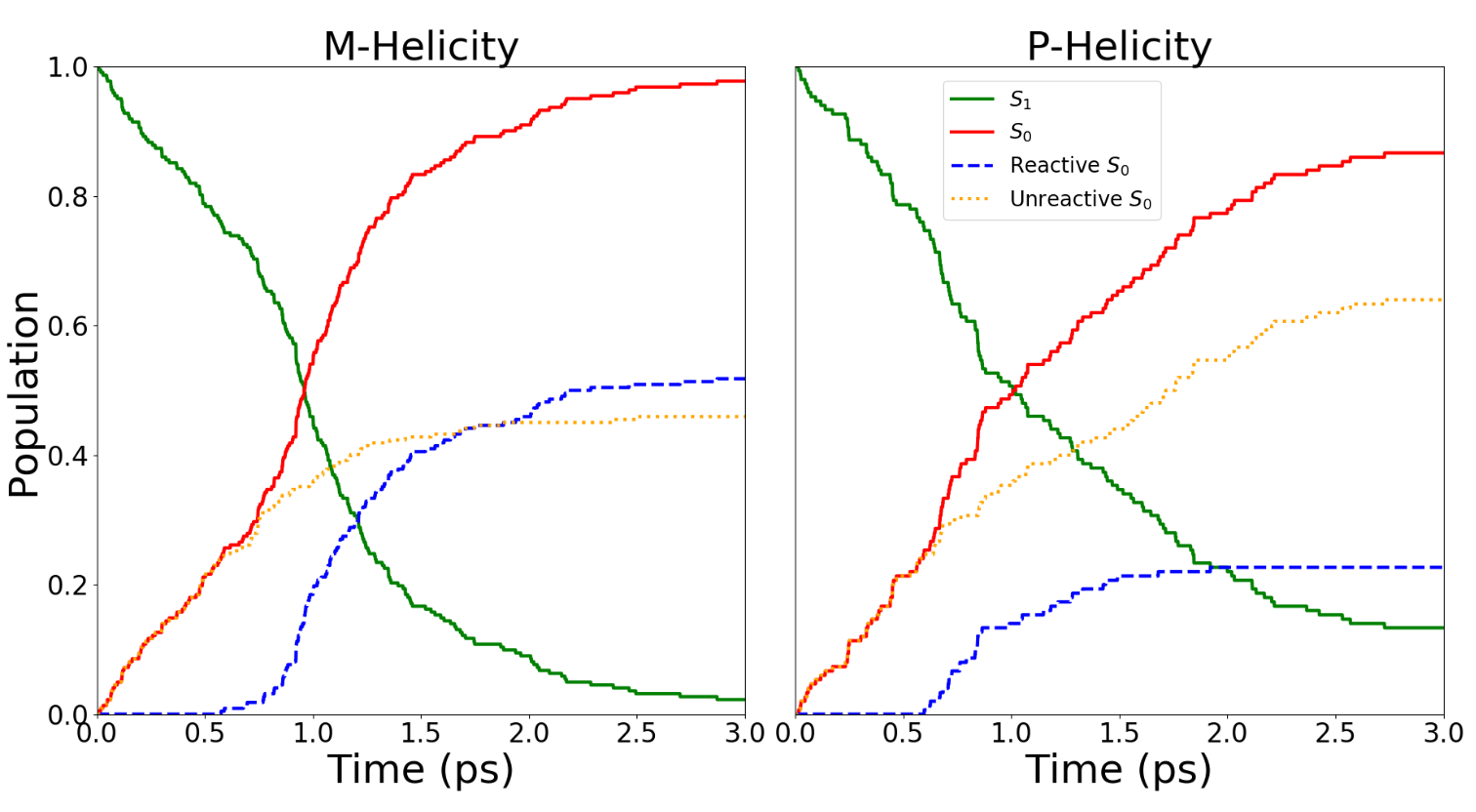}
    \caption{Population of each electronic state as a function of time for M- and P-like conformers. Color code: $S_1$ in green solid line, $S_0$ in red solid line, $S_0$ with successful isomerization in blue dashed line and $S_0$ with unsuccessful isomerization in orange dotted line.}
    \label{fig:decay}
\end{figure}
\begin{figure}
    \centering
    \includegraphics[width=\textwidth]{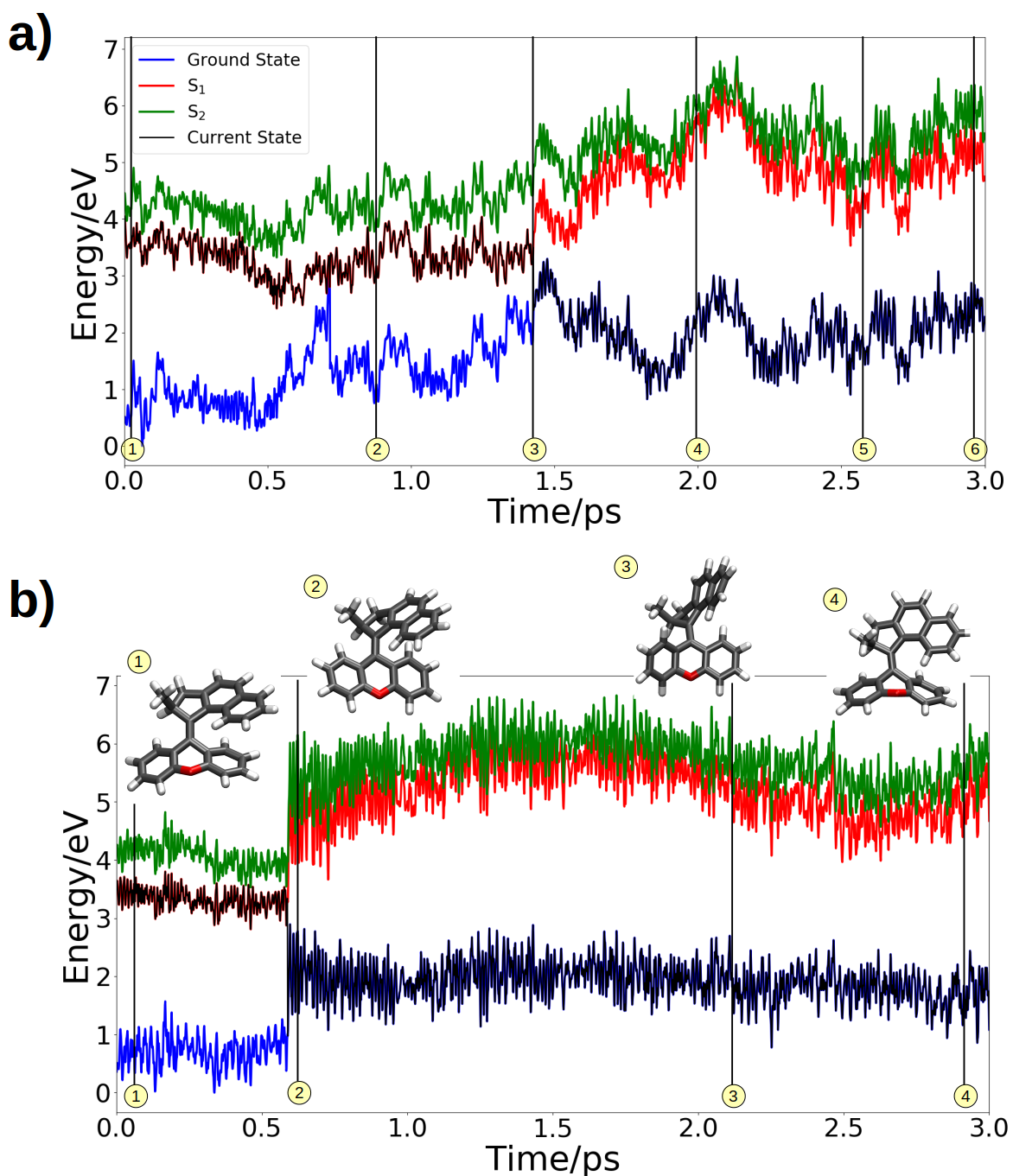}
    \caption{Potential energy of the $S_2$, $S_1$ and $S_0$ electronic states along a) a reactive trajectory with helix inversion, b) an unreactive trajectory. Structures displayed in Figure \ref{fig:dihe_time_m} are marked and labeled for the reactive trajectory. The unreactive trajectory corresponds to the one depicted in Figure \ref{fig:dihe_time_m}, panels b and d in yellow, and structures are labeled.}
    \label{fig:pes}
\end{figure}

Excited state lifetimes were determined using a monoexponential decay model. We find that M-conformers decay to the ground state faster than P-conformers, with excited state lifetimes of $\tau=0.93$ ps and $\tau=1.31$ ps, respectively (Figure \ref{fig:decay}). Furthermore, the excited state lifetimes for molecules that underwent photoisomerization were $\tau=1.21$ ps for M-helicity and $\tau=1.34$ ps for P-helicity conformers. 
Thus, for both conformers, average lifetimes are larger for the reactive channel (E-Z isomerization) than for the unreactive channel (no E-Z isomerization). 
Inspecting the population changes (Figure \ref{fig:decay}) more closely, we note that for both conformers,
trajectories relaxing to S$_0$ before 0.5 ps are all unreactive. 
Thus, there seems to be an induction period of about 0.5 ps that is needed for trajectories to successfully isomerize. 
Inspecting the time evolution of $\alpha$ of reactive (Figures \ref{fig:dihe_time_m}a and \ref{fig:dihe_time_p}a) unreactive (Figures \ref{fig:dihe_time_m}b and \ref{fig:dihe_time_p}b) trajectories, we see that reactive trajectories start to reach values close to 90$^\circ$ (for M helicity) and --90$^\circ$ (for P helicity) at about 0.5 ps, while still being in the S$_1$ state.  Unreactive trajectories, in contrast, have decayed to S$_0$ to a large fraction within the first 0.5~ps, before reaching 90$^\circ$ (for M) and --90$^\circ$ (for P). However, a few exceptions exist, where unreactive trajectories for both M and P conformers persist in the excited state beyond 0.5~ps.
We further notice that when inspecting a density representation of the time evolution of $\alpha$ (Figure S6, SI), that for M-conformers the average of $\alpha$ changes more drastically in the first 0.5 ps in reactive trajectories than in unreactive trajectories. For P-conformers, the density representation (Figure S7, SI) reveals that starting structures of unreactive trajectories (panel b) exhibit less negative average $\alpha$-values than starting structures for reactive trajectories (panel a).

\begin{figure}
    \centering
    \includegraphics[width=0.85\linewidth]{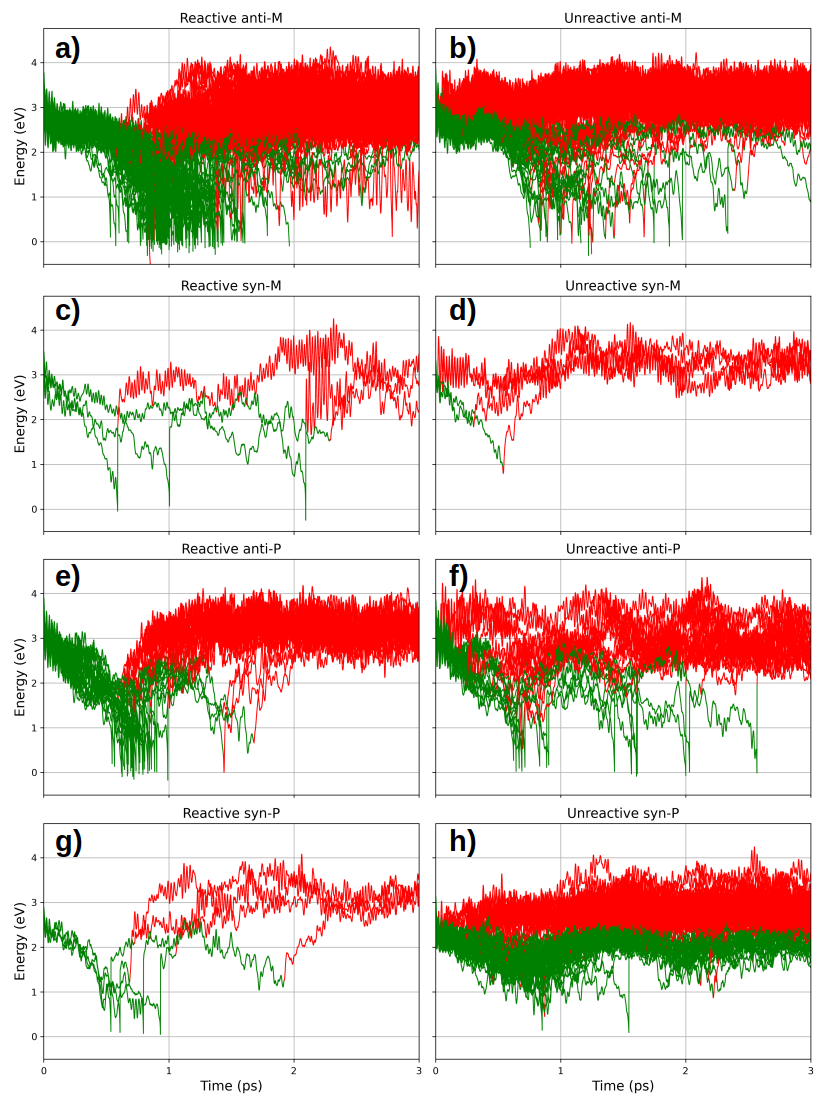}
    \caption{S$_0$ -- S$_1$ energy gap along the trajectories of each conformer. Left panels show reactive trajectories; right panels show unreactive trajectories. Green: trajectory is in S$_1$; red: trajectory is in  S$_0$.}
    \label{fig:energy_gap}
\end{figure}

These findings indicate that premature decay to S$_0$, before $\approx$0.5 ps and before reaching a value of $\pm 90$$^\circ$, prevents trajectories from isomerizing.
Since an $\alpha$ angle of $\pm 90$$^\circ$ is associated with a S$_1$--S$_0$ conical intersection, 
this is also reflected in the time evolution of the potential energies (Figure \ref{fig:pes}) and the S$_1$--S$_0$ energy gap (Figure \ref{fig:energy_gap}): a large fraction of the unreactive trajectories (example in Figure \ref{fig:pes}b) decay to the ground state before reaching a conical intersection, while reactive trajectories (example in Figure \ref{fig:pes}a) persist long enough in S$_1$ to encounter a conical intersection, before they relax to S$_0$. 
This is further confirmed by inspecting the S$_0$-S$_1$ energy gap as a function of time 
(Figure \ref{fig:energy_gap}): for example, comparing reactive anti-M trajectories (panel a) with unreactive anti-M trajectories (panel b), we see that fewer unreactive trajectories reach energy gaps below 1 eV than reactive trajectories.
We also note that about one half (52 \%) of all reactive trajectories (Figure \ref{fig:energy_gap}, panels a, c, e, g) decay to ground state due to negative excitation energies, indicating the proximity of a conical intersection; in unreactive trajectories this occurs in only 6 \% of all trajectories (Figure \ref{fig:energy_gap}, panels b, d, f, h).

The time evolution of the NAC (Figure S16, SI) shows that reactive trajectories exhibit larger magnitudes than unreactive trajectories, which is expected because they come close to the conical intersections, whereas most unreactive trajectories 
do not reach conical intersections. These trajectories decay 
 to S$_0$, despite the low NAC magnitudes, which can occur based on the statistical nature of the TDDFT-SH approach.

%

Further inspecting the dynamics of the hot ground state after excited state decay, we observe several cases of thermal helix inversion, which can be identified by a sign change of $\beta$ (Figures \ref{fig:dihe_time_m} and \ref{fig:dihe_time_p}, panels c and d).  In the case of the reactive trajectories of the M-conformers, 9 trajectories exhibit helix inversion from the primarily formed P-conformer (positive $\beta$) to the more stable M-conformer (negative $\beta$), 
as exemplified by the black line in Figure \ref{fig:dihe_time_m}c and the snapshot structures \circled{1}--\circled{6}: the primarily formed isomerized structure after isomerization (structure \circled{2} in Figure \ref{fig:dihe_time_m}) has P-helicity, whereas at later times (structures \circled{4}--\circled{6}) the molecules adopts M-helicity. Also, the unreactive M-helical trajectories exhibit a few cases (7) of at least temporary thermal helix inversion, forming the less stable P-conformer, as indicated by a change from negative $\beta$ to positive $\beta$ (Figure \ref{fig:dihe_time_m}d). In the case of the reactive trajectories of P-conformers, we only see one trajectory (Figure \ref{fig:dihe_time_p}c, black line) that changes from the initially formed M-conformer (Figure \ref{fig:dihe_time_p}, structure \circled{2}) to the less stable P-conformer (Figure \ref{fig:dihe_time_p}, structure \circled{5}). For the unreactive trajectories of the P-conformers, however, we observe 10 cases where the stable M-conformer is formed after decay to the ground state, as indicated by a change of $\beta$ from positive to negative (Figure \ref{fig:dihe_time_p}d).
In summary, we see that thermal helix inversion can occur in the hot ground state within 3 ps upon photoexcitation in all cases. However, changes from P to M are generally more likely to occur, as expected from stability considerations. Thus, the net effect of thermal helix inversion in the hot ground state favors unidirectional rotation.

\subsection{Wavelength-dependent Product Quantum Yields}
\begin{figure}
    \centering
    \includegraphics[width=\textwidth]{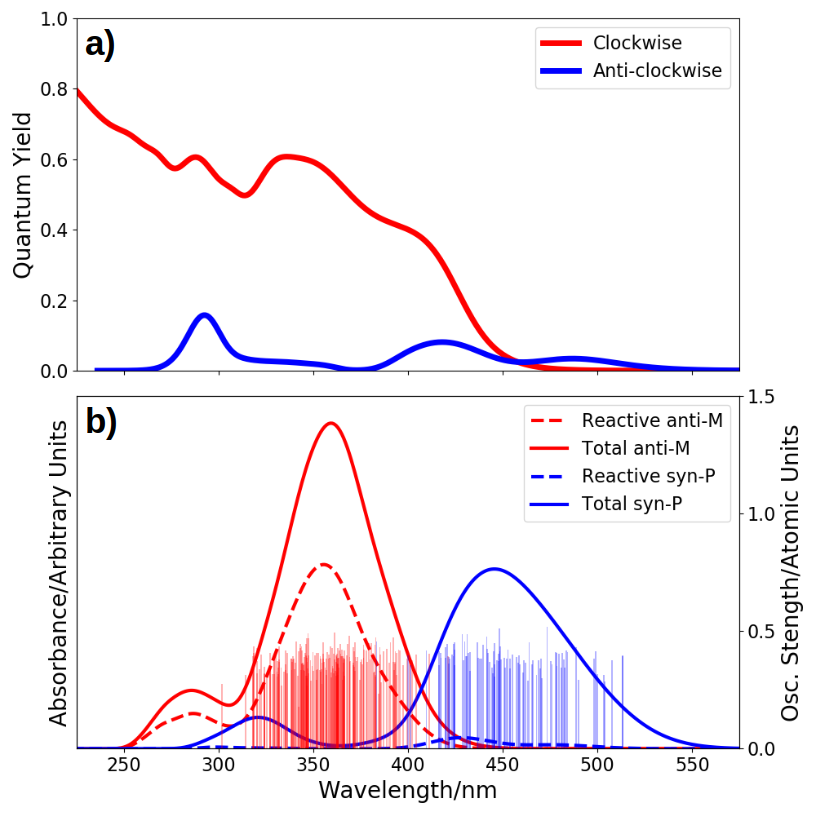}
    \caption{a) Wavelength-dependent product quantum yield for anti-M and syn-P. b) Normalized absorption spectra of the anti-M (syn-P) conformer and the one representing just the isomerized structures in dashed line. Vertical sticks represent the excitation energy and oscillator strength for the transition $S_0 \xrightarrow{} S_1$ of the initial structures.}
    \label{fig:yield}
\end{figure}

We computed the wavelength-dependent product quantum yield for the isomerization of anti-M and syn-P conformers (Figure \ref{fig:yield}a), based on Equation \ref{eq1}. The syn-M and anti-P conformers were ignored due to their negligible population at thermal equilibrium compared to anti-M and syn-P, respectively (Table \ref{tbl:pop}).
Focusing on the wavelength range between 320 to 520 nm, where sufficient sampling trajectories are present, we notice that with the exception of the long wavelength region above 460 nm, the anti-M conformer exhibits a significantly higher isomerization quantum yield than the syn-P conformer.
For anti-M, the average absorption spectrum of initial structures of reactive trajectories (Figure \ref{fig:yield}b, red, dashed) appears to be very similar to the total absorption spectrum (red, solid).
Therefore, the maximum of clockwise isomerization quantum yield at $\approx$ 340 nm (Figure \ref{fig:yield}a, red) is located close to the maximum absorption band at of the M-conformer at 360 nm
(Figure \ref{fig:yield}b, red). At this point, the probability of isomerization reaches approximately $60\%$. 
In contrast, the anti-clockwise isomerization quantum yield of syn-P exhibits two maxima red- and blue-shifted with respect to its maximum absorption band at 450 nm. The red-shifted peak reaches about 5 \%, whereas the blue-shifted peak reaches about 10 \%.
Although the quantum yield of syn-P is nearly negligible at the peak of anti-M’s major absorption band, excitation at around 375 nm results in a quantum yield of zero for syn-P, while for anti-M it remains around 50 \%. We conclude that excitation at 375 nm, would therefore 
 maximize clockwise rotation by selectively targeting anti-M conformers. 
 Nonetheless, at thermal equilibrium, the population of syn-P is approximately 300 times lower than that of anti-M (Table \ref{tbl:pop}). As a result, excitation at the peak wavelength of anti-M’s major absorption band will almost guarantee clockwise rotation.

\section{Conclusion}
In this work, we simulated the photodynamics of four conformers of a second-generation molecular rotor -- anti-M, anti-P, syn-M, and syn-P -- to investigate its ultrafast photoisomerization mechanism.

Our TDDFT-SH calculations 
confirm that the M- and P-conformers rotate in opposite directions, as previously hypothesized:\cite{unidirectionality} M-helical conformers exclusively rotate clockwise, while P-helical conformers rotate anti-clockwise.
Although our calculations suggest that the anti-M conformer is the only one significantly populated at thermal equilibrium, 
for non-equilibrium situations, which in practice could occur at high photon influx, all conformers are possibly populated. Therefore, it is advantageous if anti-clockwise rotation could be suppressed in order to maximize unidirectional rotation.
Our results predict a significantly higher clockwise isomerization quantum yield for the anti-M than anti-clockwise isomerization quantum yield for the syn-P, which has the greatest stability among the P-conformers. Moreover, at around 375 nm there is no probability of isomerization of syn-P, while for anti-M the probability is around $50\%$, coinciding with the major absorption band of the latter. Thus we predict that excitation at 375 nm maximizes unidirectional rotation.

Furthermore, our simulations reveal that thermally induced helix inversion 
can occur in the hot ground state directly after excited state decay.
While this has been observed in both directions, it generally favors unidirectional motion.
Interestingly, non-reactive decay of P-conformers accelerates helix inversion to form the more stable M-conformer.

However, our simulations completely neglected solvent effects, which may influence both the thermal helix inversion rates and the isomerization quantum yields due to solvent viscosity. Since our results demonstrate that premature excited-state decay negatively impacts isomerization quantum yields, it is reasonable to expect that increased solvent viscosity would further reduce the quantum yield by hindering trajectories from reaching the conical intersection.  

Regarding helix inversion in the hot ground state, solvent effects are also likely to decrease inversion rates. Acting as a heat bath, the solvent would dissipate nuclear kinetic energy generated during excited-state decay, making it more difficult to overcome the energetic barriers of the inversion process.  

Nevertheless, our study provides a general strategy for evaluating the wavelength dependence of isomerization quantum yields, offering a means to optimize unidirectional rotation in CPNX.

\subsection{Data availability}
The data supporting this article are included as part of the ESI.
\subsection{Conflicts of interest}
There are no conflicts to declare.
\begin{acknowledgement}
Research reported in this publication was supported by the National Institute of General Medical Sciences of the National Institutes of Health (NIH) under award numbers 1 R16GM149410-01. The content is solely the responsibility of the authors and does not necessarily represent the official views of the NIH. We acknowledge technical support from the Division of Information Technology of CSULB.
\end{acknowledgement}


\bibliography{papers}

\end{document}